\documentclass[prd,twocolumn]{revtex4}
\usepackage{graphicx}
\usepackage{amssymb}
\usepackage{amsmath}
\usepackage{amssymb}
\usepackage{pstricks}
\usepackage{epsf}
\usepackage{epsfig,graphicx,amsfonts,amssymb}

\begin{document}

\title{Do Observations Favour Galileon Over Quintessence?}
\author{Md.~Wali~Hossain\footnote{wali@ctp-jamia.res.in}, Anjan~A~ Sen\footnote{aasen@jmi.ac.in}}
\affiliation{Centre for Theoretical Physics, Jamia Millia Islamia, New Delhi-110025, India}

\date{\today}

\begin{abstract}
We study the Galileon scalar field model arising as a decoupling limit of the Dvali-Gababdaze-Porrati (DGP) construction for the late time acceleration of the universe. The model has one extra Galileon correction term over and above the standard kinetic and potential energy terms for a canonical quintessence field. We aim to study whether the current observational data can distinguish between this Galileon field and the quintessence field. Our study shows the remarkable result that for potentials like linear, square or exponential, the data prefers the Galileon model over quintessence field. It confirms that the observable universe demands the inclusion of higher derivative Galileon corrections in the standard quintessence scalar field models.
\end{abstract}

\maketitle

\date{\today}

\maketitle

Cosmological observations\cite{obs1} \cite{cmb} \cite{wmap} indicate that our present universe is going through an accelerated expanding phase. The standard lore is that an unknown form of energy, called the dark energy\cite{sami}, is responsible for driving the universe into such a late time accelerating phase. However, the nature of dark energy is still unexplained and remains a challange for particle physicists and cosmologists alike. So far, the simplest candidate for dark energy has been the cosmological constant which although is allowed by all observational data, yet it is plagued by  acute problems like fine tuning and cosmic coincidence \cite{derev}. However, current observational data can also accommodate a time varying vacuum energy. Infact, quintessence\cite{quint} ( a scalar field  which slow rolls at present energy scale)  was proposed as a candidate for dark energy to provide a dynamical solution to the cosmological constant problem. 

Recently a large scale modification has been proposed which can explain the late time acceleration at the cosmological scale. This involves an effective scalar field $\pi$ dubbed as "Galileon" \cite{galileon} as its lagrangian respects the shift symmetry in the Minkowski background: $\pi \rightarrow \pi + c$ and $\partial_{\mu}\pi \rightarrow \partial_{\mu}\pi+ b_{\mu}$ where $c$ and $b_{\mu}$ are constants. Such a field can arise in the decoupling limit of the DGP model \cite{dgp}. The lagrangian for such field can usually contain three terms: one linear in $\pi$, one contains the usual kinetic term for a canonical scalar field and the third one contains term like $(\nabla \pi)^2 \Box\pi$. This third term in particular is related to the decoupling limit of DGP model \cite{decoup}.  One can add another two terms involving higher derivatives in such a way that the final equation of motion for the $\pi$ field is still second order \cite{covariant}. This set up is theoretically appealing due to the absence of negative energy instability as well as due to  the absence of curvature singularity. The model gives rise to late time acceleration of the universe \cite{lategalileon} and at the same time is consistent in astrophysical context through Vainshtein mechanism \cite{vain}.

Although theoretically Galileon model is more exciting \cite{excit} than the standard quintessence scenario, the question is whether current observations can distinguish these two scenario. We aim to study precisely this question. Our goal is to see whether observations prefer the Galileon field over the standard quintessence  scenario. 

We consider the action for the Galileon  field in the lowest nontrivial order keeping upto the third order term in the lagrangian ( This arises in the decoupling limit of DGP model). We also keep a general potential term $V(\pi)$ in the action.

\begin{equation}
S=\int d^4x\sqrt{-g}\Bigl [\frac{M^2_{\rm{pl}}}{2} R -  \frac{1}{2}(\nabla \pi)^2\Bigl(1+\frac{\alpha}{M^3}\Box \pi\Bigr) - V(\pi) \Bigr] + \mathcal{S}_m
\label{1.1}
\end{equation}
where $M_{\rm{pl}}^2=8\pi G$ is the reduced Planck mass. $\alpha$ is a  dimensionless constant; for $\alpha=0$ this action reduces to that of a standard quintessence field. $V(\pi)$ is the potential for the $\pi$ field. For $V(\pi) = c_{1}\pi$, it is the usual third order action for the Galileon field. ${\cal S}_{m}$ is the action for the matter field. $M$ is a constant of mass dimesnion one; by a redefinition of the parameter $\alpha$, we can fix $M=M_{\rm{pl}}$.  

\begin{figure*}[t]
\centerline{\epsfxsize=6.5truein\epsfbox{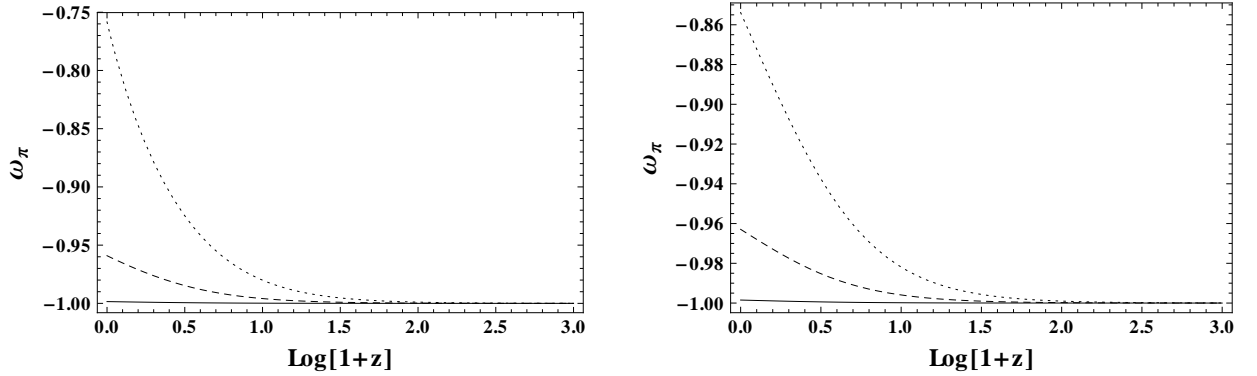}}
\caption{Evolution of the equation of state $\omega_{\pi}$ for the $\pi$ field as function of redshift. From top to bottom, $\lambda_{i} = 1, 0.5, 0.1$. $\Omega_{m0} =0.3$, $\epsilon_{i} = 15$. The left one is for linear Potential and the right one is for exponential potential}
\end{figure*}

Variation of the action (\ref{1.1}) with respect to the metric tensor $g_{\mu\nu}$ and assuming a flat Friedman-Robertson-Walker (FRW) spacetime with scale factor $a(t)$, we get the Einstein's equations:




\begin{align}
3M_{\rm{pl}}^2H^2 &=\rho_m+\frac{\dot{\pi}^2}{2}\Bigl(1-6\frac{\alpha}{M_{\rm{pl}}^{3}} H\dot{\pi}\Bigr)+V{(\pi)}\,,\\
M_{pl}^2(2\dot H + 3H^2)&=-\frac{\dot{\pi}^2}{2}\Bigl(1+2\frac{\alpha}{M_{\rm{pl}}^{3}}\ddot{\pi}\Bigr)+V(\pi)
\end{align}

Varying the action (\ref{1.1}) w.r.t the field $\pi$, we get the equation of motion for the field $\pi$ as
\begin{align}
3H\dot{\pi}+\ddot{\pi}-3\frac{\alpha}{M_{\rm{pl}}^{3}} \dot{\pi}\Bigl(3H^2\dot{\pi}+\dot{H}\dot{\pi}+2H\ddot{\pi}\Bigr)+ V'(\pi)=0
\end{align}

These above equations are supplemented by the matter conservation equation given by:
\begin{equation}
\dot\rho_m+3H\rho_m=0.
\end{equation}

Let us introduce the following dimensionless quantities

\begin{align}
x&=\frac{\dot{\pi}}{\sqrt{6}H M_{\rm{pl}}}\,,\quad y=\frac{\sqrt{V}}{\sqrt{3} H M_{\rm{pl}}}\\
\epsilon &=-6\frac{\alpha}{M_{\rm{pl}}^3}H\dot \pi\,, \quad \lambda=-M_{\rm{pl}}\frac{V'}{V}
\end{align}

Then we have the autonomous system of equations:

\begin{widetext}
\begin{align}
x'&=\frac{3 x^3 \left(2+5 \epsilon +\epsilon ^2\right)-3 x \left(2-\epsilon +y^2 (2+3 \epsilon )\right)+2 \sqrt{6} y^2 \lambda -\sqrt{6} x^2 y^2 \epsilon  \lambda }{4+4 \epsilon +x^2 \epsilon ^2},\\
y'&=-\frac{y \left(12 \left(-1+y^2\right) (1+\epsilon )-6 x^2 \left(2+4 \epsilon +\epsilon ^2\right)+\sqrt{6} x^3 \epsilon ^2 \lambda +2 \sqrt{6} x \left(2+\left(2+y^2\right) \epsilon \right) \lambda \right)}{8+8 \epsilon +2 x^2 \epsilon ^2},\\
\epsilon' &=-\frac{\epsilon  \left(-3 x \left(-3+y^2\right) (2+\epsilon )+3 x^3 \left(2+3 \epsilon +\epsilon ^2\right)-2 \sqrt{6} y^2 \lambda -\sqrt{6} x^2 y^2 \epsilon  \lambda \right)}{x \left(4+4 \epsilon +x^2 \epsilon ^2\right)},\\
\lambda' &=\sqrt{6}x\lambda^2(1-\Gamma),
\end{align}
\end{widetext}

\noindent
with $\Gamma=\frac{VV_{,\pi\pi}}{V_{,\pi}^2}$ and $\Omega_m=1-x^2 (1+\epsilon)-y^2$.  For $\epsilon=0$ we recover the autonomous system for the standard quintessence scenario \cite{scherrer&sen}. 

\noindent
The equation of state for the $\pi$ field is given by:
\begin{align}
\omega_{\pi}=\frac{-12 y^2 (1+\epsilon )+3 x^2 \left(4+8 \epsilon +\epsilon ^2\right)-2 \sqrt{6} x y^2 \epsilon  \lambda }{3 \left(4+4 \epsilon +x^2 \epsilon ^2\right) \left(y^2+x^2 (1+\epsilon )\right)}.
\end{align}

\begin{figure*}[t!]
\centerline{\epsfxsize=6.5truein\epsfbox{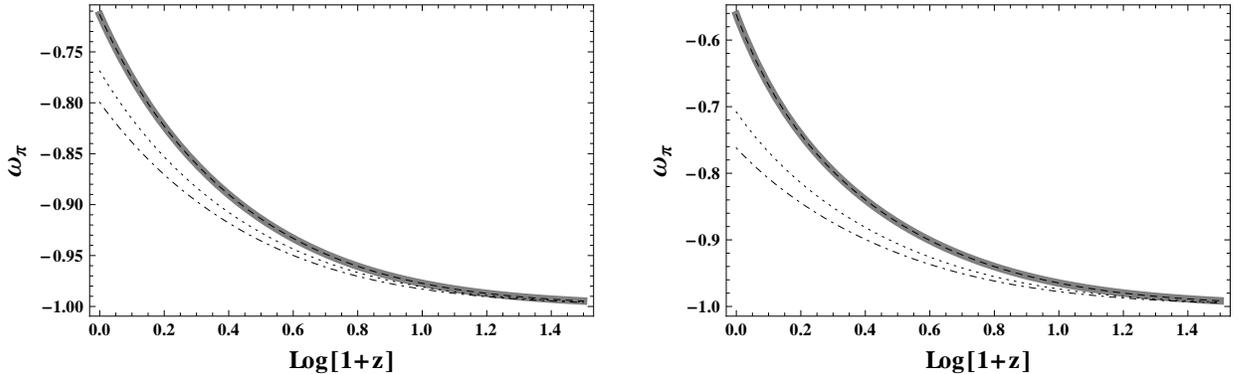}}
\caption{Evolution of the equation of state $\omega_{\pi}$ for the $\pi$ field as function of redshift. Dashed, Dotted and Dash-Dotted lines represent $\epsilon_{i} = 0, 20, 40$ whereas solid line represent the quintessence field. Potential is chosen to be linear one. The left one is for $\Omega_{m0} = 0.3$ and the right one is for $\Omega_{m0} = 0.24$}.
\end{figure*}

In our system of equations (9)-(12), we have four variables e.g $x, y, \epsilon$ and $\lambda$. The variable $\epsilon$ sets the relative strength of the Galileon correction term over the standard kinetic energy term for a quintessence field. To solve this system, we need to specify the initial values for these four variables. We set our initial condition at the decoupling time $z=1000$ and evolve the system from then onwards till the present epoch $z=0$. 

\noindent
Initially the $\pi$ field is nearly frozen due to large hubble damping. This sets $x_{i}$ to be small initially which keeps equation of state $\omega_{\pi}$ very close to $-1$. We vary $x_{i}$ between $0.01$ and $10^{-7}$ and check that initial value of $\omega_{\pi}$  does not vary much and remains extremely close to $-1$.  The initial value for $\epsilon$, $\epsilon_{i}$, sets the initial strength of the Galileon correction over the the standard quintessence term. If this value is zero initially, it remains zero throughout the history of the universe and the evolution is same as the standard quintessence scenario. Hence the value of $\epsilon_{i}$ determines the deviation from the quintessence scenario and the effect of the Galileon correction. So in our study we keep this parameter $\epsilon_{i}$ as a model parameter. The parameter $\lambda$ is the slope of the potential and its initial value determines the slope at which the field starts rolling initially. For smaller values of $\lambda_{i}$, the $\pi$ field behaves very close to the  cosmological constant (C.C). This is irrespective of the form of potential or whether the $\pi$ field behaves as quintessence or Galileon. As one increases $\lambda_{i}$, the $\pi$ field starts behaving differently from  C.C. This is shown in Figure 1 where we plot the equation of state for the $\pi$ field, $\omega_{\pi}$, for two potentials (Exponential and Linear) for different values of $\lambda_{i}$. We plot this for Galileon field assuming $\epsilon_{i} = 15$. 

Our aim is to distinguish between the quintessence and the Galileon field and hence we concentrate in the region where they deviate substantially from C.C. For this we set $\lambda_{i} =1$ in our subsequent calculations which ensures that the $\pi$ field deviates from the C.C behaviour at present.

With the above choices for $x_{i}, \epsilon_{i}$ and $\lambda_{i}$, the initial value of the variable $y$, $y_{i}$,  is related to the present day matter density $\Omega_{m0}$. The parameter $\Gamma$ controls the shape of the potential. In our study, we consider four different forms for the potential, e.g linear, squared, exponential and inverse squared. These are the most well studied forms of potential for the scalar field model of dark energy.

In Figure 2, we show the behaviour of the equation of state $\omega_{\pi}$ as a function of redshift for different values of $\epsilon_{i}$. We choose two values for $\Omega_{m0}$ e.g 0.24 and 0.3. We show it for the linear potential, but the overall behaviour remains the same for other potentials. From this figure, it is apparent that for $\epsilon_{i} =0$ initially, the $\pi$ behaves exactly same as quintessence. As one increases the $\epsilon_{i}$, the two models starts deviating from each other. This deviation is slightly higher for smaller value of $\Omega_{m0}$.  Hence constraining $\epsilon_{i}$ by observational data is crucial to distinguish between quintessence and Galileon field.  

In the rest of the paper, we concentrate on this issue. For this, we consider various observational data currently available.

We consider the Supernovae Type Ia observation which is one of the direct probes for late time acceleration.We have utilized the Union2 compilation of the dataset which comprises of 557 datapoints \cite{union2}. It measures the apparent brightness of the Supernovae as observed by us which is related to the luminosity distance $d_{L}(z)$ 
defined  as 
\begin{equation}
d_{L}(z) = (1+z)\int_0^z\frac{dz^{\prime}}{H(z^{\prime})}\end{equation}

With this we construct the distance modulus `$\mu$' which is experimentally measured:
\begin{equation}
\mu = m-M = 5\log\frac{d_{L}}{Mpc}+25.
\end{equation}
Here m and M are the apparent and absolute magnitudes of the Supernovae which are logarithmic measure of flux and luminosity respectively. 

Another observational probe that has been widely used in recent times to constrain dark energy models is related to the data from the Baryon Acoustic Oscillations measurements \cite{bao} by the large scale galaxy survey. In this case, one needs to calculate the parameter $D_{v}$ which is related to the angular diameter distance as follows
\begin{equation}
D_{v} = \left[\frac{z_{BAO}}{H(z_{BAO})}\left(\int_0^{z_{BAO}}\frac{dz}{H(z)}\right)^2\right] ^{1/3}.
\end{equation}
For BAO measurements we calculate the ratio 

{\Large$\frac{D_{v}(z = 0.35)}{D_{v}(z = 0.20)} $}.
 \vspace{1mm}
This ratio is a relatively model independent quantity and has a measured value $1.736 \pm 0.065$. 

\begin{figure*}[t]
\centerline{\epsfxsize=4.2truein\epsfbox{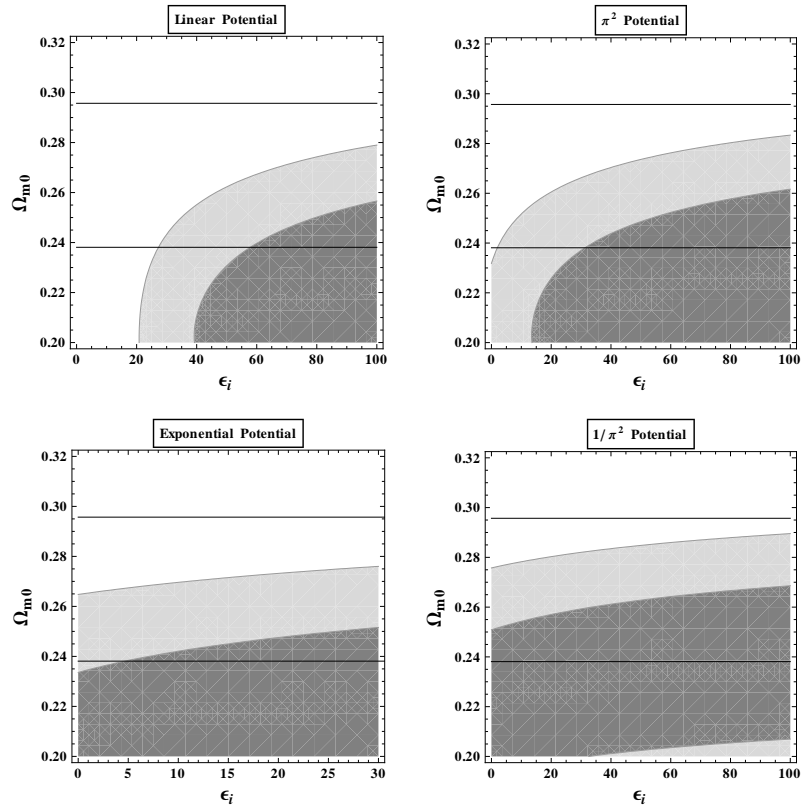}}
\caption{Confidence contour in $\epsilon_{i}-\Omega_{m0}$ plane for different potentials using SN+BAO+HUBBLE data.
The dark region represent the confidence region at $68\%$ confidence level whereas the dark+light region represents the region at $95\%$ confidence level.
}
\end{figure*}
Next we use  new determinations of the cosmic expansion history from red-envelope galaxies. Stern et al. \cite{stern} have obtained a high-quality spectra with the Keck-LRIS spectrograph of red-envelope galaxies in 24 galaxy clusters in the redshift range $0.2 < z < 1.0$. They complemented these Keck spectra with high-quality, publicly available archival spectra from the SPICES and VVDS surveys. With this, they presented 12 measurements of the Hubble parameter H(z) at different redshift. The measurement at $z=0$ was from HST Key project \cite{H0}.At this point we define the normalised hubble parameter as $h(z) = \frac{H(z)}{H_{0}}$ and utilise it to derive the value of new h(z) and its deviation. 

Using all these observational data, our aim is to constrain the parameter $\epsilon_{i}$.  We do not make any assumption on range of $\epsilon_{i}$ which may arise due to the underlying theory. We simply want to see what is allowed by observational data.

\noindent
The result is shown in Figure 3.  Here we show the confidence contours in the $\epsilon_{i}-\Omega_{m0}$ parameter space for various potentials.  We also show the bound on $\Omega_{m0}$ as obtained by WMAP 7 observation which is given by $\Omega_{m0} = 0.2669\pm 0.0288$. The most striking result is that for both linear potential and square potentials, the quintessence case ($\epsilon_{i} = 0$) is ruled out by more than $95\%$ ( $2\sigma$) confidence level. For the exponential potential, the quintessence case is ruled out at $68\%$ ($1\sigma$) confidence level. Only for the inverse squared potential both the quintessence and the Galileon behaviours are allowed  both at $1\sigma$ and $2\sigma$ confidence level. We should stress that the linear potential is the most simple and well studied case for Galileon field.  Hence for this simple potential as well as for other potentials, quintessence field is clearly distinguishable from the Galileon field by current observational data. This result also confirms the fact that the observational data actually prefers the Galileon correction term in the scalar field lagrangian. 

To conclude, the inclusion of Galileon correction term in the standard canonical scalar field lagrangian is well motivated and elegant. Usually one can include five terms in a scalar field lagrangian which respect Galileon symmetry atleast in the flat case and also result second order equation of motion avoiding the appearances of instability. Our aim is to see whether such inclusion of higher derivative terms is also observationally interesting. In other words whether cosmological observations prefer such terms over the standard canonical kinetic term.  We stress that our goal is not to compare Galileon model with $\Lambda$CDM which is consistent with all the current obervations. Hence we consider the parameter region where the $\pi$ field behaves differently from C.C.

\noindent
We restricts ourselves to the simple case of Galileon model which contains three terms in the lagrangian and which arises in the decoupling limit of the DGP model. By fitting this model with observational data, we show that data indeed prefers this Galileon correction over the standard canonical kinetic term for most of the potentials including the most simple linear potential. This, in our knowledge, is the first result confirming that current observations demand the inclusion of Galileon correction in the scalar field lagrangian. This makes the Galileon model even more interesting.

\vskip 2mm
The authors are grateful to M. Sami for his useful comments and suggestions during various stages of this work. MWH acknowledges the discussions with R. Gannouji and Amna Ali regarding various aspects of Galileon models. The author AAS acknowledges the Department of Science and Techonology, Govt of India for reserach funding through project no: SR/S2/HEP-43/2009. MWH acknowleges the funding from CSIR, Govt of India.


\begin{thebibliography}{99}

\bibitem{obs1}A. G. Riess et al., Astron. J. {\bf 116}, 1009 (1998);
S. Perlmutter et al., Astrophys. J. {\bf 517}, 565 (1999); J. L.
Tonry et al., Astrophys. J. {\bf 594}, 1 (2003).

\bibitem{cmb}A. Melchiorri et al., Astrophys. J. Lett. {\bf 536},
L63 (2000); A. E. Lange et al., Phys. Rev. D {\bf 63}, 042001
(2001); A. H. Jaffe et al., Phys. Rev. Lett. {\bf 86}, 3475
(2001); C. B. Netterfield et al., Astrophys. J. {\bf 571}, 604
(2002); N. W. Halverson et al., Astrophys. J. {\bf 568}, 38
(2002).

\bibitem{wmap}S. Bridle, O. Lahab, J. P. Ostriker and P. J. Steinhardt,
                 Science {\bf 299}, 1532 (2003); C. Bennett et al.,
Astrophys. J. Suppl. Ser. {\bf 148}, 1 (2003); G. Hinshaw et al.,
Astrophys. J. Suppl. Ser. {\bf 148}, 135 (2003); A. Kogut et al.,
Astrophys. J. Suppl. Ser. {\bf 148}, 161 (2003);
              D. N. Spergel et al., Astrophys. J. Suppl. Ser. {\bf 148}, 175 (2003).

\bibitem{sami} E. J. Copeland, M. Sami and S. Tsujikawa, Int. J. Mod. Phys. D {\bf 15}, 1753 (2006);
Miao Li, Xiao-Dong Li, Shuang Wang, 
arXiv:1103.5870 

\bibitem{derev}
S.~Weinberg,
Rev.\ Mod.\ Phys.\  {\bf 61}, 1 (1989);\\
V.~Sahni and A.~A.~Starobinsky,
 Int.\ J.\ Mod.\ Phys.\  D {\bf 9}, 373 (2000)
  [astro-ph/9904398];\\
S.~M.~Carroll,
 Living Rev.\ Rel.\  {\bf 4}, 1 (2001)
  [astro-ph/0004075];\\
P.~J.~E.~Peebles and B.~Ratra,
 Rev.\ Mod.\ Phys.\  {\bf 75}, 559 (2003)
  [astro-ph/0207347];\\
T.~Padmanabhan,
Phys.\ Rept.\  {\bf 380}, 235 (2003)
  [hep-th/0212290]

\bibitem{quint}
 P.~J.~E.~Peebles and  B.~Ratra, \apj {\bf 325}, L17 (1988);
C.~Wetterich, Nucl.\ Phys.\ B {\bf 302}, 668 (1988); M. S. Turner
and M. White, Phys. Rev. D {\bf{56}}, 4439 (1997); R. R. Caldwell,
R. Dave and P. J. Steinhardt, Phys. Rev. Lett. {\bf{80}}, 1582
(1998);  I.~Zlatev, L.~M.~Wang and P.~J.~Steinhardt, Phys.\ Rev.\
Lett.\ {\bf 82}, 896 (1999).

\bibitem{galileon} A.~Nicolis, R.~Rattazzi and E.~Trincherini, Phys.~Rev.~D, {\bf 79}, 064036 (2009).

\bibitem{dgp}G.~R.~Dvali, G.~Gabaddze and M.~Porrati, Phys.~Lett.~B, {\bf 485}, 208, (2000).

\bibitem{decoup}M.~A.~Luty, M.~Porrati and R.~Rattazzi,  JHEP,  {\bf 09}, 029 (2003);
A.~Nicolis and R. ~Rattazzi,  JHEP, {\bf 06}, 059 (2004). 


\bibitem{covariant}C.~Deffayet, G.~Esposito-Farese and A.~Vikman, Phys.~Rev.~D, {\bf 79}, 084003 (2009);
C.~Deffayet, S.~Deser and G.~Esposito-Farese, Phys.~Rev.~D, {\bf 80}, 064015 (2009);
N.~Chow and J.~Khoury, Phys.~Rev.D, {\bf 80}, 024037 (2009);


\bibitem{lategalileon}A.~Ali, R.~Gannouji and M.~Sami, Phys.~Rev.~D, {\bf 82}, 103015 (2010);
 R.~Gannouji and M.~Sami, Phys.~Rev.~D, {\bf 82}, 024011 (2010);
 A.~De.~Felice and S. Tsujikawa, Phys.~Rev.~Lett., {\bf 105}, 111301 (2010);
 S.~A.~Appleby and E.~Linder, arXiv:1112.1981;
 E.~Linder, arXiv:1201.5127.
 
 \bibitem{vain}A.~I.~Vainshtein, Phys.~Lett.~B, {\bf 39}, 393 (1972).
 
 \bibitem{excit}
 G.~Goon, K.~Hinterbichler and M.~Trodden, JCAP, {\bf 12}, 004 (2011);
 C.~de~Rham and L. Heisenberg, Phys.~Rev.~D, {\bf 84}, 043503 (2011);
 M.~Trodden and K.~Hinterbichler, Class.~Quant.~Grav., {\bf 28}, 204003 (2011);
 C.~Burrage and C.~de.~Rham and L.~Heisenberg, JCAP, {\bf 05}, 025 (2011).
 
 \bibitem{scherrer&sen}R.~J.Scherrer and A.~A.~Sen, Phys.~Rev.~D, {\bf 77}, 083515 (2008).
 
 \bibitem{union2}R.~Amanullah, et al. Astrophys.~J., {\bf 716}, 712 (2010).
 
 \bibitem{bao}W.~J.~Percival, et al, Mon.~Not.~Roy.~Astron.~Soc., {\bf 401}, 2148 (2010).
 
 \bibitem{stern} D.~Stern, R.~Jimenez, L.~Verde, M.~Kamionkowski and S.~Adam,  JCAP, {\bf 1002}, 008 (2010).
 
 \bibitem{H0}W.~L.~Freedman, et al. Astrophys.~J., {\bf 553}, 47 (2001).
 
 \end{thebibliography}
\end{document}